\documentclass[12pt]{article}
\usepackage{graphicx}
\usepackage{wrapfig}
\usepackage{amssymb}
\usepackage{cite}

\usepackage[a4paper,top=1.75cm,bottom=1.75cm,left=2.5cm,right=2.5cm,bindingoffset=0mm]{geometry}

\usepackage{xcolor}
\usepackage{hyperref}

\hypersetup{
	colorlinks,
	citecolor=blue,
	linkcolor=blue,
	urlcolor=blue
}

\usepackage{tikz}
\usepackage{tikz-3dplot}

\usepackage{enumitem}
\usepackage{hyperref}
\usepackage{cite}
\usepackage{booktabs}
\usepackage{tabularx}

\usepackage{pifont}

\usetikzlibrary{shapes, arrows}
\usetikzlibrary{decorations.pathreplacing}
\usetikzlibrary{positioning, calc}
\tikzstyle{fitted} = [rectangle, minimum width=5cm, minimum height=1cm, text centered, draw=black, fill=red!30]
\tikzstyle{operations} = [rectangle, rounded corners, minimum width=2cm,text centered, draw=black, fill=red!30]
\tikzstyle{roundtext} = [rectangle, rounded corners, minimum width=2cm, minimum height=0.8cm, text centered, draw=black, fill=red!30]
\tikzstyle{n3py} = [rectangle, rounded corners, minimum width=3cm, minimum height=1cm, text centered, draw=black, fill=green!30]
\tikzstyle{myarrow} = [thick,->,>=stealth]
\tikzstyle{line} =[draw, -latex']
\tikzstyle{decision} = [diamond, draw, fill=red!20, text width=7.5em, text centered,  inner sep=0pt, minimum height=2em, aspect=4]
\tikzstyle{cloud} = [draw, ellipse,fill=green!20, minimum height=2em]
\tikzstyle{inout} = [rectangle, draw, fill=green!20, text width=9.5em, text centered, rounded corners, minimum height=2em, minimum width=10em]
\tikzstyle{block}=[rectangle, draw, fill=blue!20, text width=9.5em, 
text centered, rounded corners, minimum height=2em, 
minimum width=10em]

\definecolor{darkgreen}{rgb}{0.0, 0.5, 0.13}

\newcommand\pubdate{\today}

\def\Title#1{\begin{center} {\Large #1 } \end{center}}
\def\Author#1{\begin{center}{ \sc #1} \end{center}}
\def\Address#1{\begin{center}{ \it #1} \end{center}}

\newcommand\pubblock{\rightline{\begin{tabular}{l}  \\ 
         \pubdate  \end{tabular}}}
\newenvironment{Abstract}{\begin{quotation}  }{\end{quotation}}
\newenvironment{Presented}{\begin{quotation} \begin{center} 
             PRESENTED AT\end{center}\bigskip 
      \begin{center}\begin{large}}{\end{large}\end{center} \end{quotation}}

\begin{document}
\begin{titlepage}
 \pubblock
\vfill
\Title{\Huge Towards an integrated determination of \\[+0.3cm] proton, deuteron and nuclear PDFs}
\vfill
\Author{Tanjona R. Rabemananjara}
\Address{Department of Physics and Astronomy, Vrije Universiteit, NL-1081 HV Amsterdam\\[0.1cm]
  	Nikhef Theory Group, Science Park 105, 1098 XG Amsterdam, The Netherlands\\[0.1cm]}
\vfill
\begin{Abstract}
  We present progress towards a unified framework enabling the simultaneous  determination of
  the parton distribution functions (PDFs) of the proton, deuteron, and nuclei up to lead $(^{208}\rm{Pb})$.
Our approach is based on the 
integration of the fitting framework underlying the nNNPDF3.0 determination of nuclear PDFs
into that adopted for the NNPDF4.0 global analysis of proton PDFs.
Our work paves the
way toward a full integrated global analysis of non-perturbative QCD -- a key ingredient for the 
exploitation of the scientific potential of
present and future nuclear and particle physics facilities such as the Electron-Ion Collider 
(EIC).
\end{Abstract}
\vfill
\begin{Presented}
DIS2023: XXX International Workshop on Deep-Inelastic Scattering and
Related Subjects, \\
Michigan State University, USA, 27-31 March 2023 \\
     \includegraphics[width=9cm]{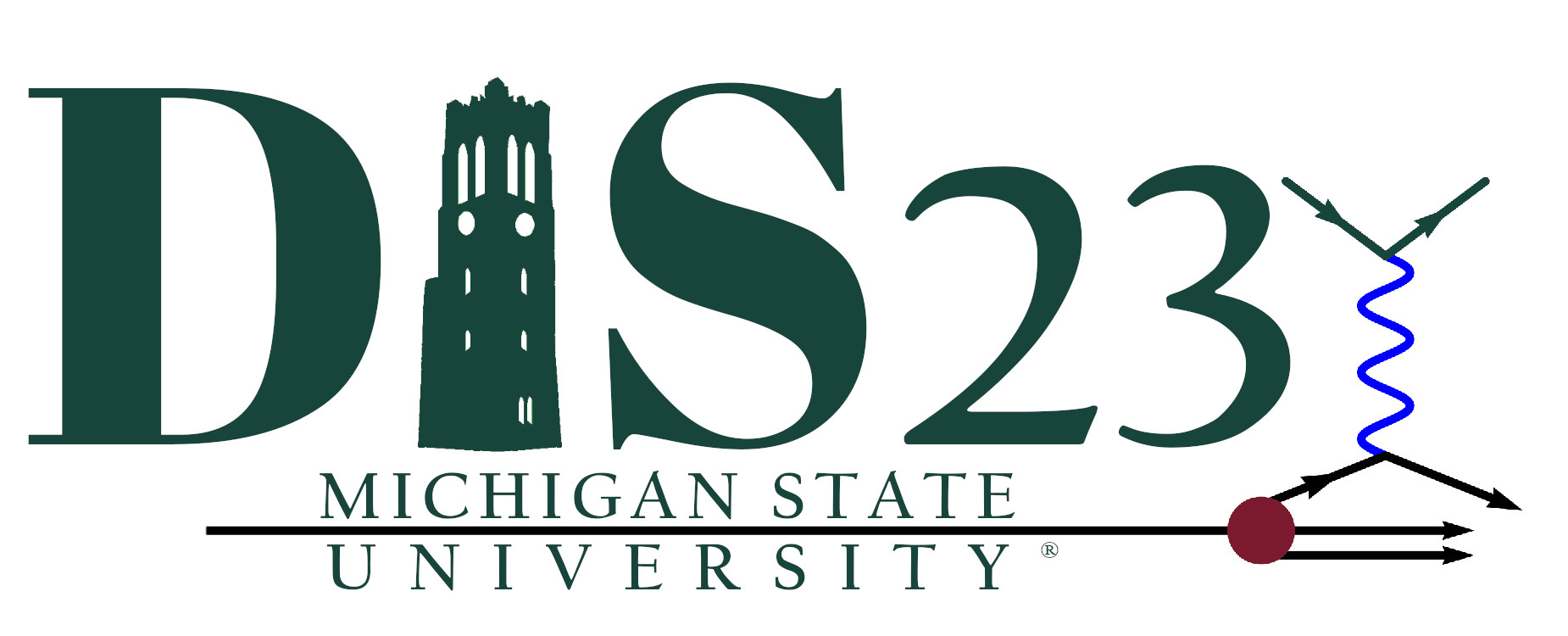}
\end{Presented}
\vfill
\end{titlepage}


\paragraph{Introduction.}
Future measurements at the recently approved Electron-Ion
Collider (EIC)~\cite{AbdulKhalek:2022hcn,AbdulKhalek:2021gbh} and at the High-Luminosity 
Large Hadron Collider (HL-LHC)~\cite{Azzi:2019yne} will perform key measurements that can pin down the
dynamics of Quantum Chromodynamics (QCD).
By colliding
(polarized) beams with proton or deuteron or heavier nuclei for a range of center-of-mass 
energies, these experiments will be fundamental in understanding how partons are distributed 
in momentum and position spaces within a proton and how the parton distribution of nucleon 
bound within nuclei are modified w.r.t. their free-nucleon counterparts.

In order to fully
interpret the precision-level measurements offered by such experiments, the theoretical
determination of both the unpolarized proton and nuclear parton distribution functions, 
henceforth referred to as (n)PDFs, needs to be improved.
While the determination of the free-proton PDFs has seen quite an advancement in recent 
years~\cite{Ethier:2020way} 
-- owing mostly to the availability of  a broad
experimental dataset and theoretical improvements in higher-order calculations
-- the current status in the determination of nuclear PDFs is somewhat less advanced.
In addition,
albeit both proton and nuclear PDFs are simultaneously constrained by datasets 
where one of the targets or projectiles
is not a free-state proton, their extractions are usually performed in 
a separate manner.

On one hand, it has been well understood that measurements involving deuteron and heavier 
nuclear targets play a significant role in disentangling the proton's quark and antiquark
distributions, and in separating the up and down PDFs for large momentum fractions in which
searches for physics beyond the Standard Model (BSM) are relevant~\cite{Ball:2022qtp}.
In the past, different
approaches have been adopted to account for nuclear corrections in proton PDF fits, each with
its own motivations and limitations. The approach adopted in the NNPDF 
methodology~\cite{NNPDF:2021njg, NNPDF:2021uiq}
consists in adding the uncertainties due to nuclear effects as an extra contribution to the
theory covariance matrix~\cite{Ball:2018lag, Ball:2018twp, Ball:2020xqw}.

On the other hand, all hadronic datasets included in the determination of nuclear PDFs involve
a free-proton in the initial state. Most nuclear PDF sets are therefore determined assuming a 
fixed proton baseline which can be considered as a theoretical bias whose effect is difficult 
to estimate. In the nNNPDF methodology~\cite{AbdulKhalek:2019mzd, AbdulKhalek:2020yuc, AbdulKhalek:2022fyi}, 
although the $A=1$ dependence 
(with $A$ representing the atomic mass number) is also fitted in the same footing as $A \neq 1$ 
by means of a neural network, a fixed proton baseline is still required to enforce the $A \to 1$ 
limit, therefore introducing a potential bias in that the back-reaction
of the fitted data on the proton PDf is ignored.
Such a constraint is imposed at the level of 
the $\chi^2$ as a penalty by means of a Lagrange multiplier.

Motivated by the need for a consistent and concurrent extractions of the unpolarized
proton, deuteron, and heavier nuclear PDFs, we present here
work in progress towards an ``integrated fit'' (see also~\cite{Khalek:2021ulf} for a first attempt) in 
which the atomic mass number dependence is smoothly parametrized from $A=1$ to $A=208$
by fitting simultaneously to proton, deuteron, and heavier nuclear datasets,
removing the need for imposing a boundary condition for the $A=1$ limit.
A similar idea was applied to PDFs and fragmentation functions
by the JAM collaboration in~\cite{Moffat:2021dji,Ethier:2020way}.
Our approach is
based on the integration of the fitting framework underlying the nNNPDF3.0 determination 
of nuclear PDFs into that adopted for the NNPDF4.0 global analysis of proton PDFs~\cite{NNPDF:2021uiq}. 
As a first attempt to apply our methodology, we perform fits in which only deep inelastic
scattering 
(DIS) processes are included using NNLO QCD calculations with perturbative charm.

Here we briefly describe the integrated fitting
framework, emphasizing on the its main differences w.r.t the (n)NNPDF approaches. 
We then study the stability of the integrated fits based on the NNPDF4.0 default hyperparameters 
from which all the subsequent results are derived. We assess the impacts of the integrated
fitting methodology on the proton and nuclear PDFs by comparing the results with the reference
(n)NNPDF determination. Conclusions are drawn in the last section.


\paragraph{Methodology.}
As mentioned above, the integrated fitting methodology incorporates the nuclear
PDF parametrization from nNNPDF3.0 into the determination of free-proton PDF using the NNPDF4.0
methodology. The reason for this is twofold. First, albeit the nNNPDF3.0 is also based on 
deterministic minimization algorithm it does not provide some of the advanced machine learning
techniques that the NNPDF4.0 possesses -- such as the automatic tuning of the hyperparameters
using the $k$-folding procedure~\cite{NNPDF:2021njg, NNPDF:2021uiq}. In addition, the NNPDF4.0 
methodology also provides
ways to carefully validate the resulting PDF uncertainties via the closure and future test
approaches. Nonetheless, the two determinations share a large number of similarities, one among which is 
the parametrization of the (n)PDFs in the evolution basis.

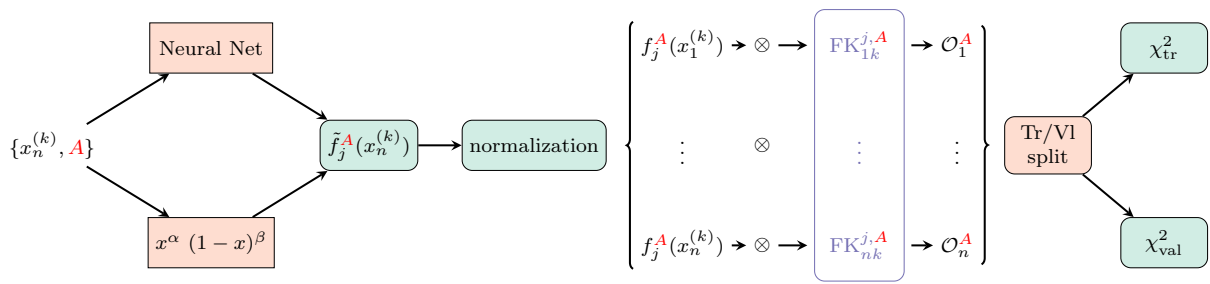
\begin{figure}[!h]
	\centering
	\resizebox{1.0\textwidth}{!}{%
		\begin{tikzpicture}[node distance = 1.0cm]\scriptsize
			\definecolor{vp1}{RGB}{102,194,165}
			\definecolor{vp2}{RGB}{252,141,98}
			\definecolor{vp3}{RGB}{117,112,179}
			\node (xinput) {$\{x_{n}^{(k)}, \textcolor{red}{A}\}$};
			
			\coordinate [right = 1.5cm of xinput] (NNghost) {};
			\node[fitted, fill=vp2!30, above = 1.0cm of NNghost, minimum width=1.7cm, minimum height=0.7cm]
			(pdf) { Neural Net};
			\node[fitted, fill=vp2!30, below = 1.0cm of NNghost, minimum width=1.7cm, minimum height=0.7cm]
			(preproc) { $x^{\alpha}$ $(1-x)^{\beta}$};
			
			\node[operations, fill=vp1!30, minimum width=1.2cm, minimum height=0.7cm, right = 1.5cm of NNghost]
			(fitbasis) {$\tilde{f}_{j}^{\textcolor{red}{A}}(x^{(k)}_n)$};
			\node[operations, fill=vp1!30, minimum width=1.2cm, minimum height=0.7cm, right = 0.6cm of fitbasis]
			(normalizer) {normalization};
			
			\node[right = 0.9cm of normalizer] (pdfdots) {\vdots};
			\node[above = 0.7cm of pdfdots]
			(pdf1) {${f}_{j}^{\textcolor{red}{A}}(x^{(k)}_1)$};
			\node[below = 0.7cm of pdfdots]
			(pdfn) {${f}_{j}^{\textcolor{red}{A}}(x^{(k)}_n)$};
			
			\node[right = 0.2cm of pdf1] (conv1) {$\otimes$};
			\node[right = 0.2cm of pdfn] (convn) {$\otimes$};
			\node at ($(conv1)!0.5!(convn)$) (convdots) {$\otimes$};
			
			\node[vp3, right = 0.6cm of conv1] (f1) {FK$_{1k}^{j, \textcolor{red}{A}}$};
			\node[vp3, right = 0.6cm of convn] (fn) {FK$_{nk}^{j, \textcolor{red}{A}}$};
			\node[vp3] at ($(f1)!0.5!(fn)$) (fd) {\vdots};
			\draw[draw=vp3, rounded corners] ($(f1.north west)+(-0.1, 0.2)$) rectangle ($(fn.south east)+(0.1,-0.2)$);
			
			\node[right = 0.5 cm of f1] (o1) {$\mathcal{O}_{1}^{\textcolor{red}{A}}$};
			\node[right = 0.5 cm of fn] (on) {$\mathcal{O}_{n}^{\textcolor{red}{A}}$};
			\node at ($(o1)!0.5!(on)$) (od) {\vdots};
			
			\node[operations, fill=vp2!30, right = 0.5cm of od, minimum width = 1.2cm, text width=1cm, minimum height=0.7cm]
			(trvl) {Tr/Vl split};
			\coordinate [right = 1.0cm of trvl] (ending) {};
			\path let \p1 = (ending), \p2 = (pdf)
			in node at (\x1,\y2) [n3py, fill=vp1!30, minimum width = 1.2cm, minimum height=0.7cm] (tr) {$\chi^{2}_{\rm tr}$};
			\path let \p1 = (ending), \p2 = (preproc)
			in node at (\x1,\y2) [n3py, fill=vp1!30, minimum width = 1.2cm, minimum height=0.7cm] (vl) {$\chi^{2}_{\rm val}$};
			
			\draw[myarrow] (xinput) -- (pdf);
			\draw[myarrow] (xinput) -- (preproc);
			\draw[myarrow] (pdf) -- (fitbasis);
			\draw[myarrow] (preproc) -- (fitbasis);
			\draw[myarrow] (fitbasis) -- (normalizer);
			
			\draw[myarrow] (pdf1) -- (conv1);
			\draw[myarrow] (pdfn) -- (convn);
			\draw[myarrow] (conv1) -- ($(f1.west)-(0.2,0.0)$) ;
			\draw[myarrow] (convn) -- ($(fn.west)-(0.2,0.0)$) ;
			\draw[myarrow] ($(f1.east)+(0.2,0.0)$) -- (o1);
			\draw[myarrow] ($(fn.east)+(0.2,0.0)$) -- (on);
			
			\draw[myarrow] (trvl) -- (tr);
			\draw[myarrow] (trvl) -- (vl);
			
			\draw[decorate, decoration={brace}, thick] (pdfn.south west) -- (pdf1.north west);
			\draw[decorate, decoration={brace},thick] (o1.north east) -- (on.south east);
		\end{tikzpicture}
	}
	\caption{A modified version of Fig.~\textcolor{blue}{3.2} from~\cite{NNPDF:2021njg} with the additional 
		$\textcolor{red}{A}$-dependence.}
	\label{fig:n3fit}
\end{figure}

The main modification to the NNPDF4.0 methodology to account for the nuclear fit consists in
parametrizing the $\textcolor{red}{A}$-dependence of the (n)PDFs during the fit. That is, the 
relation between the output of the neural network and the (n)PDFs is given by:
\begin{equation}
	xf^{\textcolor{red}{A}}_k(x, Q_0; \theta) = \eta_k^{\textcolor{red}{A}} 
	x^{1-\alpha_k^{\textcolor{red}{A}}} (1-x)^{\beta_k^{\textcolor{red}{A}}} 
	\mathrm{NN}_k^{\textcolor{red}{A}} (x, Q_0; \theta),
\end{equation}
where $k$ runs over the elements of the PDF in the evolution basis, 
$\mathrm{NN}_k^{\textcolor{red}{A}}(x; \theta)$ is the $k$-th output of the neural network, 
$\theta$ indicates the full set of neural network parameters, $\eta_k^{\textcolor{red}{A}}$
is the normalization corresponding to the $k$-th PDF, and ${\textcolor{red}{A}}$ 
represents the atomic mass number of the proton/deuteron/nucleus. Note that in principle
the preprocessing exponents $\alpha_k^{\textcolor{red}{A}}$ and $\beta_k^{\textcolor{red}{A}}$ 
should also depend on $\textcolor{red}{A}$. The way in which such a dependence on the atomic 
mass number ${\textcolor{red}{A}}$  is propagated through the fitting framework is 
illustrated in Fig.~\ref{fig:n3fit}.

In order to test the integrated framework, we focus in the present study on fits to DIS datasets
only which include measurements on proton, deuteron, and nuclear targets as comprised
in the (n)NNPDF determinations. As theory inputs, we use NNLO QCD calculations provided by the
new theory pipeline used in the NNPDF framework~\cite{Barontini:2023vmr}.
And in order to be able to compare to previous 
nNNPDF PDF releases, we take the approach in which charm is generated perturbatively. The fitting 
scale and the kinematic cuts are taken to be the same as in the default NNPDF4.0 methodology, 
refer to~\cite{NNPDF:2021njg, NNPDF:2021uiq} for more details.


\paragraph{Training stability.}
The baseline hyperparameters used in the present analysis for the integrated fit are the same as the
ones used in the default NNPDF4.0 methodology extracted from a $k$-folding hyperoptimization  
procedure. These are listed in Table~\textcolor{blue}{3.7}
of~\cite{NNPDF:2021njg} with the difference that now the maximum number of epochs is larger to account for the
longer training. Owing to the more complex two-dimensional parameter space $(x, A)$ to be fitted
\cite{Candido:2023utz} -- as opposed to the one-dimensional space relevant for a free-proton PDF determination in
which only the $x$-dependence is parametrized -- one indeed expects the integrated fit to converge slower.

Table~\ref{table:chi2} compares the fit quality between a free-proton fit determined using the default NNPDF4.0
methodology and a combined proton and nuclear fit determined using the integrated framework. Displayed
are the total experimental $\chi^2_{\rm exp}$ per data point for all the datasets entering each
determination, the average experimental $\langle \chi^2_{\rm exp} \rangle$ over the replica, and the
experimental training and validation error functions $\langle E_{\rm tr}\rangle$ and 
$\langle E_{\rm val} \rangle$.
\begin{figure}[!tb]
	\centering
	\includegraphics[width=0.496\linewidth]{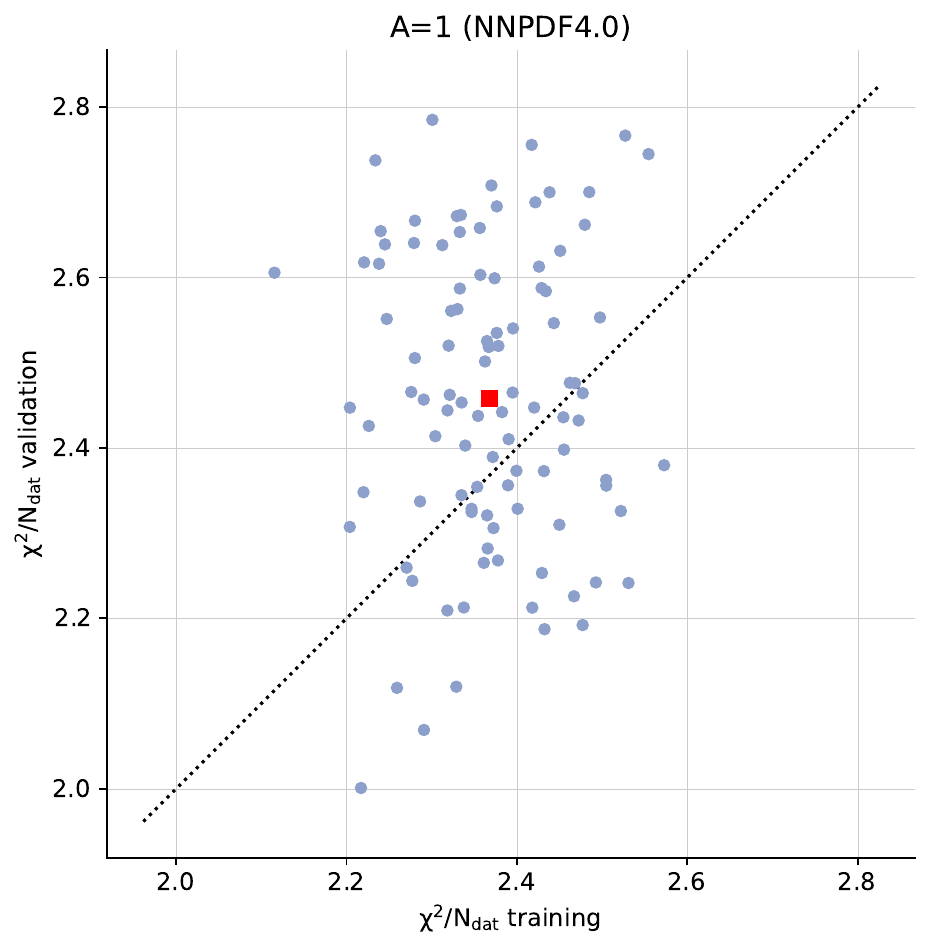}
	\includegraphics[width=0.496\linewidth]{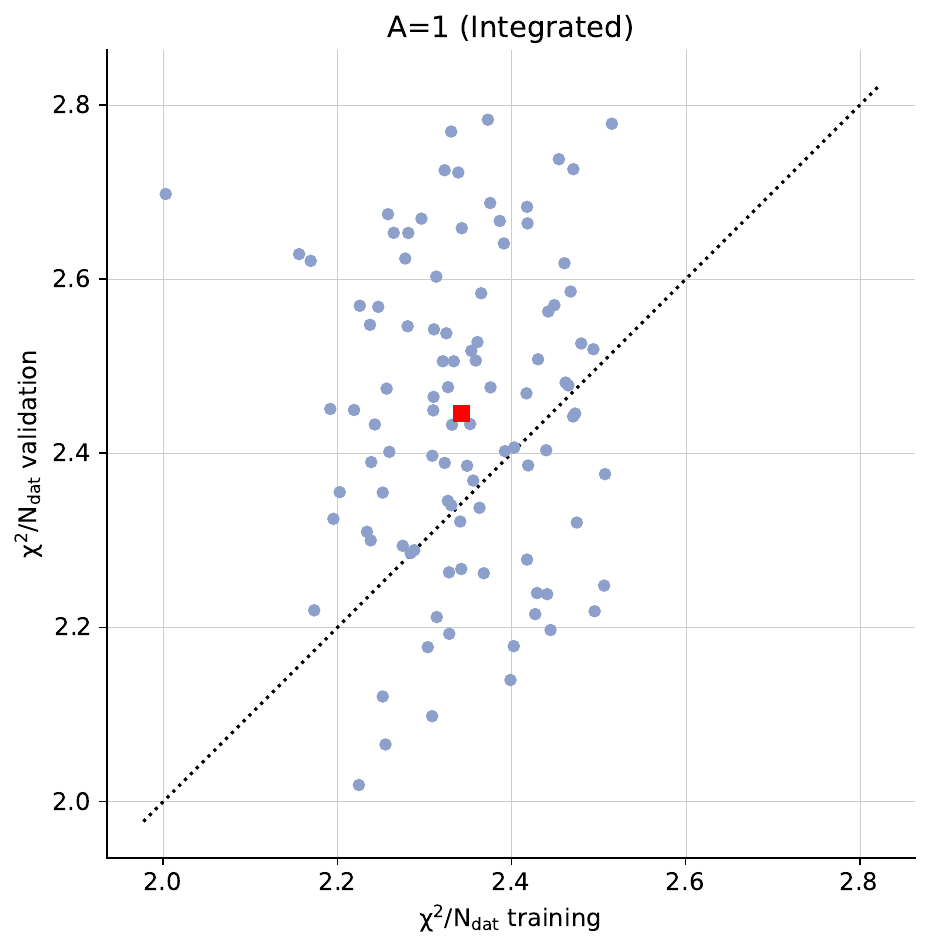}
	\caption{Distribution of the training $\chi^2$ values over the different replicas for the default
	NNPDF4.0 (left) and integrated methodologies (right). The red square represents the mean
	value over the replicas.}    
	\label{fig:chi2-distribution}
\end{figure}
\begin{figure}[!tb]
	\centering
	\includegraphics[width=0.496\linewidth]{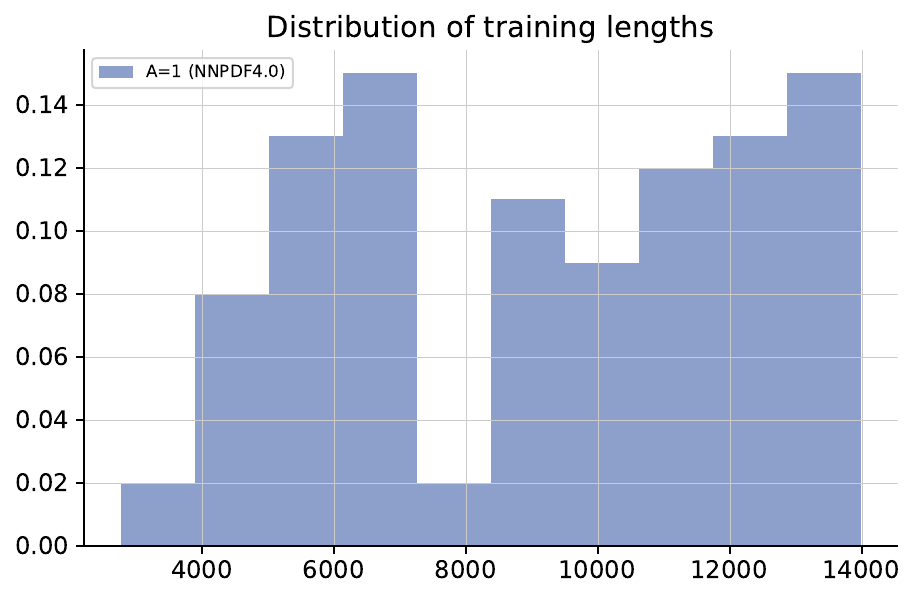}
	\includegraphics[width=0.496\linewidth]{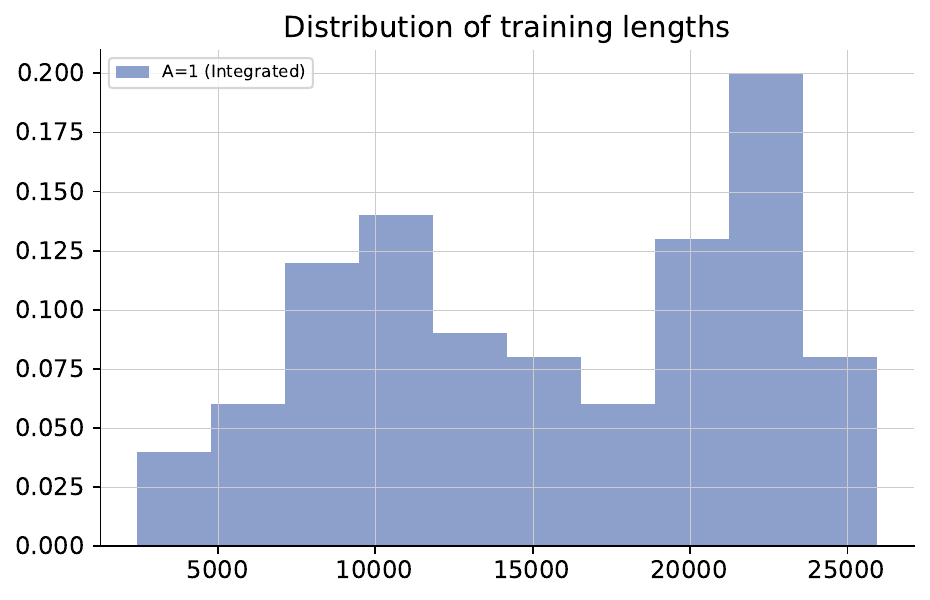}
	\caption{Distribution of the training lengths over the replicas that passed the postfit
	selection criteria for the default NNPDF4.0 (left) and integrated methodologies (right).}
	\label{fig:training-distribution}
\end{figure}
\begin{table}[!h]
	\centering
	\small
	\renewcommand{\arraystretch}{1}
	\begin{tabularx}{\textwidth}{XccXXXX}
		\toprule
		Methodology & $\chi^2_{\rm exp}$ & $N_{\rm dat}$ & $\langle \chi^2_{\rm exp} \rangle$ & $\langle E_{\rm tr}\rangle$
		& $\langle E_{\rm val} \rangle$ & Tr. Lengths \\
		\midrule
		NNPDF4.0 & 1.291 & 1788 & 1.336$\pm$0.028 & 2.328$\pm$0.087 & 2.440$\pm$0.180 & 9200$\pm$3200  \\
		\midrule
		Integrated & 1.322 & 3161 & 1.357$\pm$0.030  & 2.360$\pm$0.088  & 2.460$\pm$0.170 & 15700$\pm$6400  \\
		\bottomrule
	\end{tabularx}
	\caption{Comparison of the fit quality between the free-proton only and integrated fits.}
	\label{table:chi2}
\end{table}

From Table~\ref{table:chi2} one finds that in the integrated fit a good description of the experimental
data which includes nuclear datasets is achieved, with a total of $\chi^2_{\rm exp} = 1.322$ per data
point. Looking at the $\chi^2$-breakdown we see that the total experimental $\chi^2_{\rm exp}(A=1) = 1.192$
on the proton datasets is much smaller than the value obtained from the free-proton only fit. This might
potentially be a sign of an unbalanced fit in that the integrated fit is slightly overlearning along the 
proton direction -- which contains much more data points -- while underlearning along the nuclear directions.
Such an instability could be seen when looking at the distribution of the training $E^{(k)}_{\rm tr}$ and
validation $E^{(k)}_{\rm val}$ errors evaluated over the different MC replicas as shown in 
Fig.~\ref{fig:chi2-distribution}. In general, one expects that the values of $E^{(k)}_{\rm tr}$ are smaller
than those of $E^{(k)}_{\rm val}$, however, as opposed to the case of the free-proton fit, very few points
are located below the diagonal line. This observation is further strengthened by looking at the distribution
of training lengths as shown in Fig.~\ref{fig:training-distribution} in which one can see that not only most 
of the replicas reached the maximum number of iterations but also exhibit a bimodal distribution. This
indicates that in order to further stabilize the integrated fits, a $k$-folding hyperoptimization -- on the
combined proton, deuteron, and nuclear datasets -- is necessary to find the best combinations of 
hyperparameters. Given that this is an exhaustive task, we henceforth, present results based on the default
hyperparameters.
\begin{figure}[!tb]
	\centering
	\includegraphics[width=0.496\linewidth]{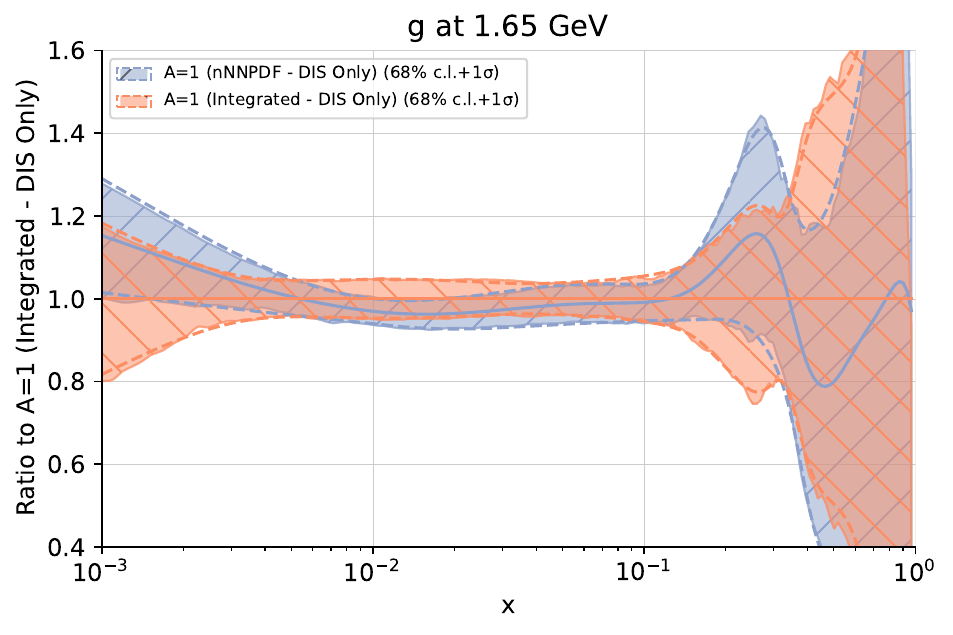}
	\includegraphics[width=0.496\linewidth]{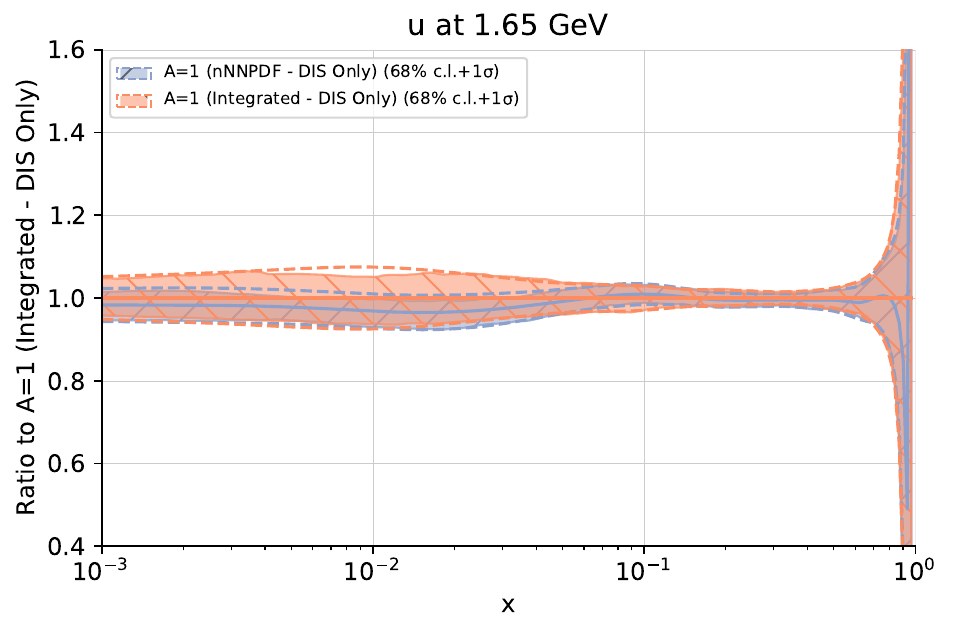}
	\caption{Comparison of the free-proton PDFs determined using the default NNPDF4.0 methodology and
		the integrated methodology in which the nuclear PDFs are also determined simultaneously. Solid and
		dashed bands correspond to 68\% c.l. and one-sigma uncertainties, respectively.}    
	\label{fig:proton-pdfs}
\end{figure}
\begin{figure}[!tb]
	\centering
	\includegraphics[width=0.496\linewidth]{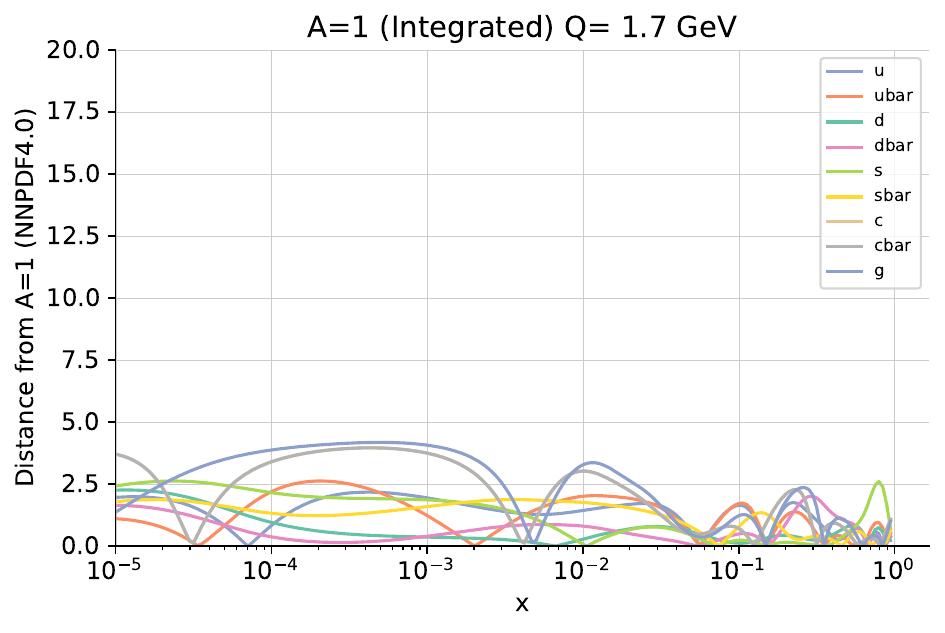}
	\includegraphics[width=0.496\linewidth]{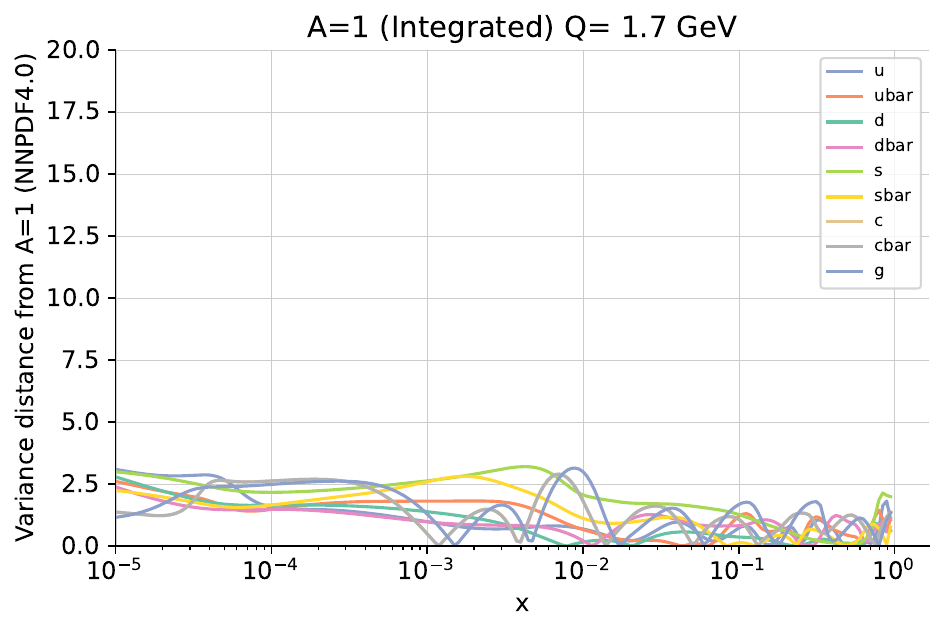}
	\caption{Distance between the central values (left) and the uncertainties (right) of the integrated 
		free-proton PDF determination w.r.t the proton PDF determined using the NNPDF4.0 methodology. 
		The results are shown for all PDF flavors at $Q^2=1.7~\rm{GeV}^2$.}
	\label{fig:proton-distance}
\end{figure}


\paragraph{Towards integrated (n)PDFs.}
%
\begin{figure}[!tb]
	\centering
	\includegraphics[width=0.496\linewidth]{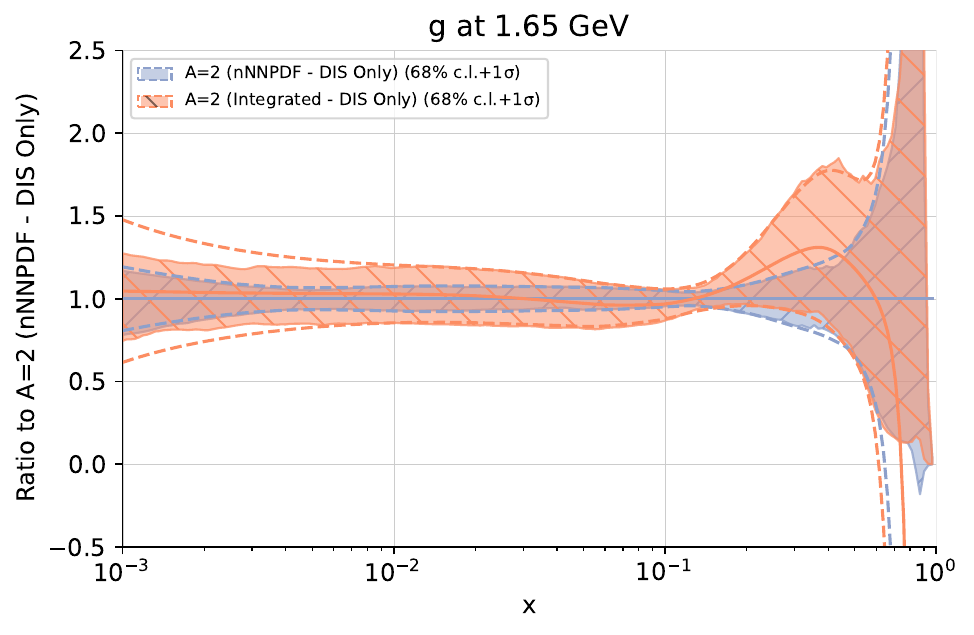}
	\includegraphics[width=0.496\linewidth]{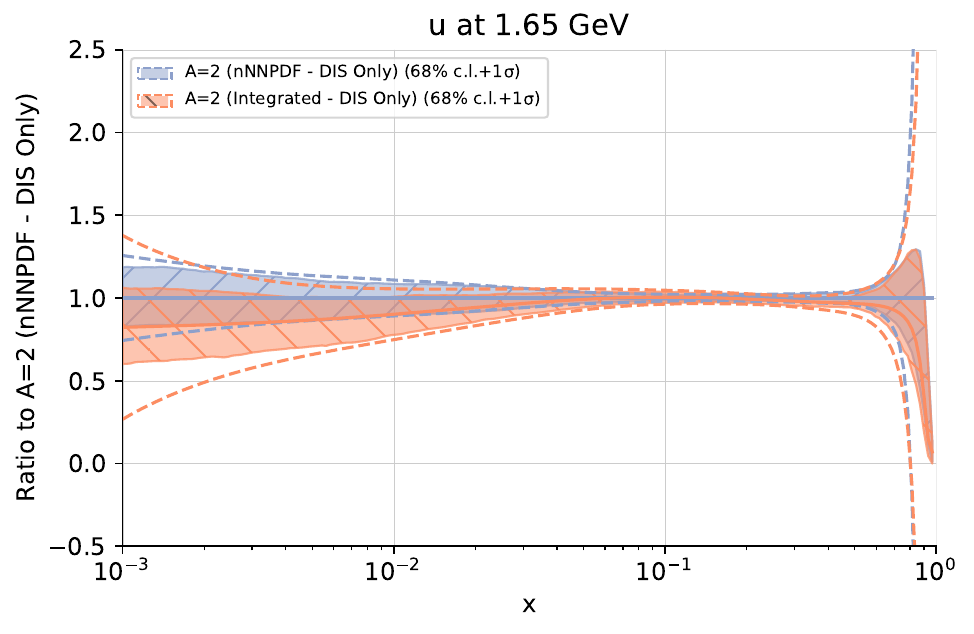}
	\includegraphics[width=0.496\linewidth]{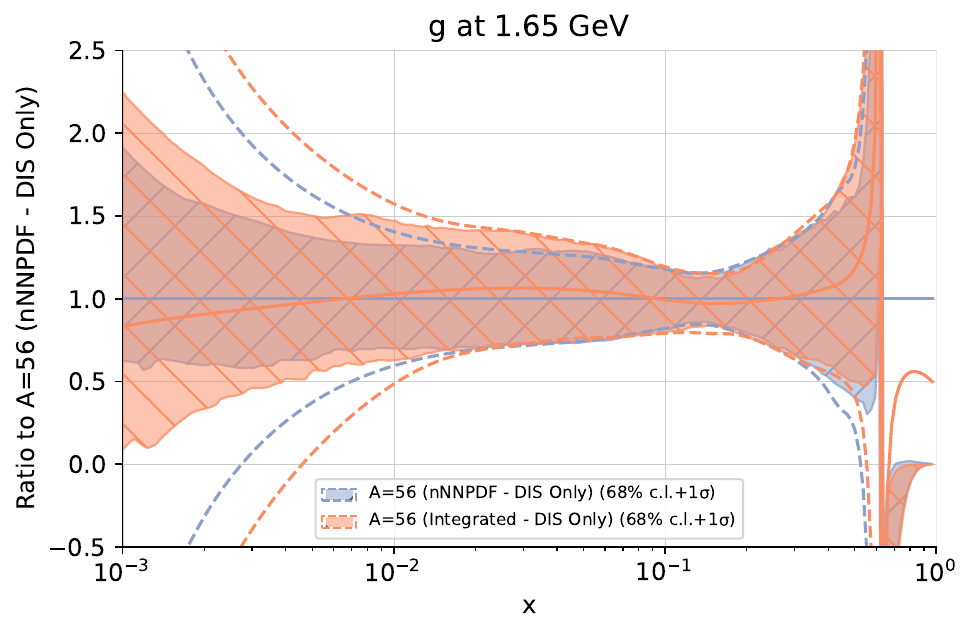}
	\includegraphics[width=0.496\linewidth]{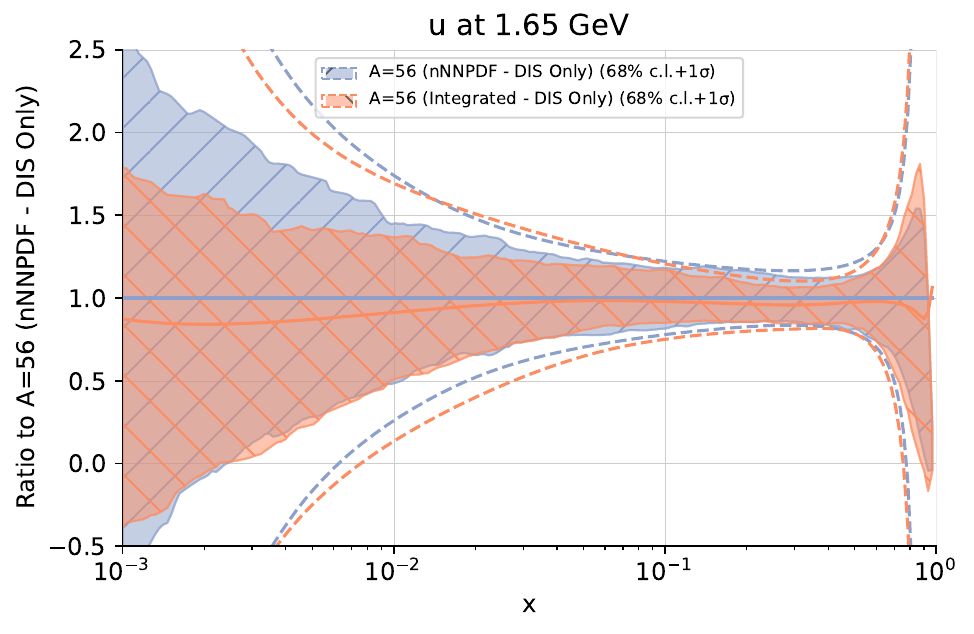}
	\includegraphics[width=0.496\linewidth]{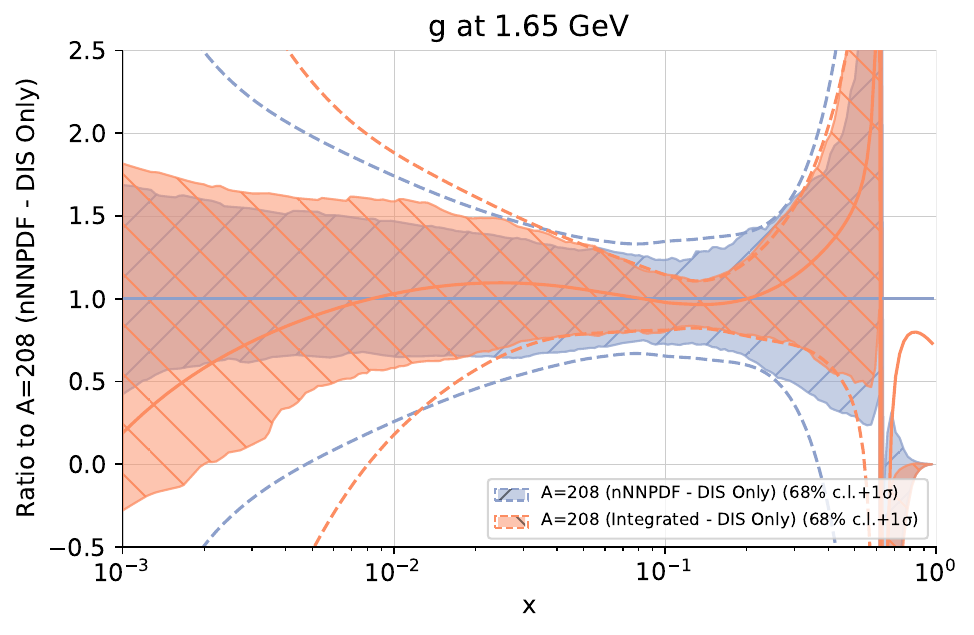}
	\includegraphics[width=0.496\linewidth]{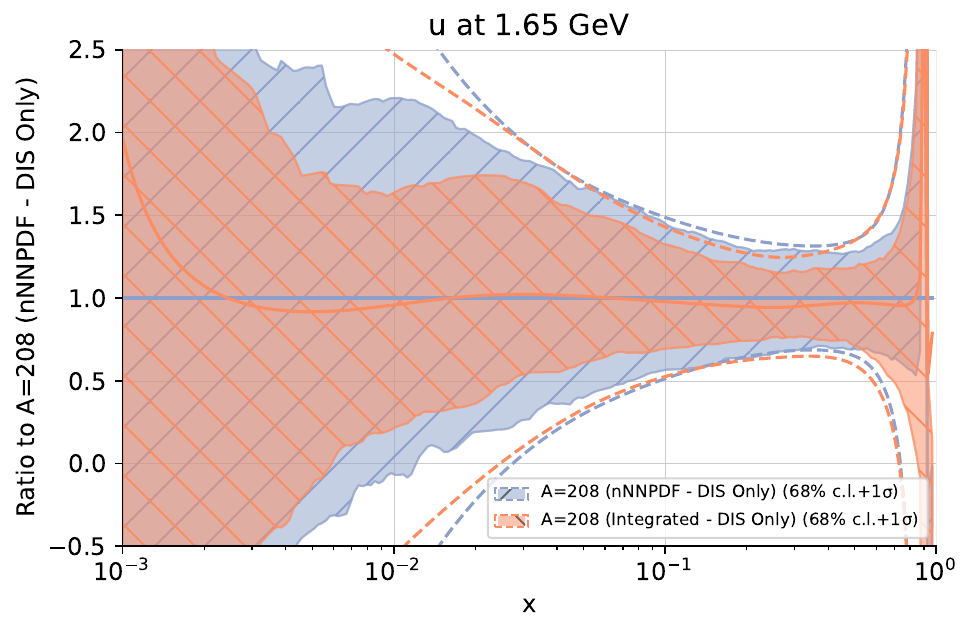}
	\caption{Comparison between the nNNPDF and integrated approaches for nuclear PDF determination. The
	results are shown for $^2\rm{D}$, $^{56}\rm{Fe}$, and $^{208}\rm{Pb}$ for the up and gluon nPDFs at
	$Q^2=1.65~\rm{GeV}^2$. The results are normalized to the central values of the PDFs determined from
	the nNNPDF approach. The solid and dashed bands represent the 68\% c.l. and one-sigma uncertainties,
	respectively.}    
	\label{fig:nuclear-pdfs}
\end{figure}
We now study the impacts of the integrated determination by comparing the resulting (n)PDFs with the proton
and nuclear PDFs determined using the NNPDF4.0 and nNNPDF methodology respectively. We recall that the
comparisons shown henceforth are for DIS-only fits with NNLO QCD theory and perturbative charm.

Let us first examine the proton-PDFs determined using the integrated approach. In Fig.~\ref{fig:proton-pdfs}
we show the comparison of the NNPDF4.0 and integrated approaches for the up and gluon PDFs as a function of $x$ 
at the fitting scale $Q_0=1.65~\rm{GeV}^2$. The results are normalized to the central values of the PDFs 
determined in the integrated approach. The solid and dashed uncertainty bands represent the 68\% confidence 
level interval and the one-sigma error, respectively. In general, there is an excellent agreement between the 
proton PDFs determined using the default NNPDF4.0 methodology and the integrated approach. Marked wiggles can 
be observed in the integrated determination which could be connected to the slight overfitting discussed in the 
previous section. The consistency between the two approaches is further supported by the distance plots shown in
Fig.~\ref{fig:proton-distance}. Displayed are the absolute and variance distance measured from the default
NNPDF4.0 determination on the flavor basis as a function of $x$ at $Q^2=1.7~\rm{GeV}^2$. As we can see, the
differences in both the central values and uncertainties are well within one sigma.

We now turn to the comparisons of the deuteron and heavier nuclear PDFs. Fig.~\ref{fig:nuclear-pdfs}
displays the comparisons between the nNNPDF and integrated approaches for the deuteron ($^2\rm{D}$), and
two heavier nuclei -- namely Iron ($^{56}\rm{Fe}$) and lead ($^{208}\rm{Pb}$). The results are shown as a
function of $x$ for the up and gluon nPDFs at $Q^2=1.65~\rm{GeV}^2$. Similar to the previous results,
the solid and dashed bands represent the 68\% c.l. and one-sigma uncertainties, respectively. All results
are normalized to the the central values of the integrated nPDFs. The two determinations are overall in
excellent agreement within the uncertainties although in general the nPDFs determined using the integrated
approach yield larger uncertainties and more fluctuations. This can be markedly seen for instance for the
gluon PDF of the deuteron. More pronounced differences are observed for both the up and gluon PDF of the
lead in which the integrated method yield much smaller error at medium- and large-$x$.

These comparisons demonstrate that albeit the issues regarding the stability of the integrated fit -- which
is directly linked to the choice of hyperparameters -- the framework is working and is capable of reproducing
the reference fits in which the proton and nuclear PDFs are separately determined.


\paragraph{Conclusions and outlook.}
We presented a framework in which the proton, deuteron, and heavy nuclear parton distributions are
simultaneously determined without the need for imposing a boundary condition to reproduce the $A=1$
limit. It is based on the integration of the framework underpinning the nNNPDF3.0 determination of the
parton distribution of nucleons bound within nuclei into that adopted in the NNPDF4.0 methodology for
the determination of free-proton PDFs.

It was shown that the (n)PDFs extracted from the integrated approach are consistent with the reference
(n)NNPDF determinations. However, using the default NNPDF4.0 set of hyperparameters did not yield
desirable results due to the sign of instability in the training. After all, the underlying nature of the neural
network architecture has drastically changed due to the additional parametrization of the atomic mass
number $A$. Therefore, a $k$-folding-based scan of the parameter space must be performed in order to 
obtain the best combination of hyperparameters.

From the physics' point of view, in order to achieve a reliable separation between quark flavors, we must
extend the experimental datasets to include hadronic processes. All the ingredients should be indeed
available to perform a global nuclear PDF fit with NNLO QCD calculations with fitted charm. Once this is
done, the next steps will be to include estimation of missing higher order 
uncertainties~\cite{NNPDF:2019vjt, NNPDF:2019ubu, Ball:2021icz}, and
to provide an approximate N3LO (n)PDF integrated determination~\cite{Hekhorn:2023gul, McGowan:2022nag}.

\paragraph{Acknowledgments.}
The author is grateful to Juan Rojo for the careful reading of the manuscript. The author also wishes
to thank the collaborators from the Netherlands eScience Center (NLeSC) for discussions during the development
of the project. This work is supported by an Accelerating Scientific Discoveries grant of the NLeSC.


\bibliographystyle{JHEP}
\bibliography{uqcd}
 
\end{document}